\documentclass{appolb}
\pdfoutput=1
\usepackage[usenames,dvipsnames]{color}
\definecolor{darkblue}{RGB}{0,0,196}
\usepackage[colorlinks=true,linkcolor=darkblue,citecolor=darkblue,urlcolor=darkblue]{hyperref}
\usepackage{graphicx}
\usepackage{makecell}
\usepackage{bm}
\usepackage{amsmath}

\begin{document}
\title{Highly-anisotropic hydrodynamics for central collisions 
\thanks{Talk presented at XI Workshop on Particle Correlations and Femtoscopy, 3-7 November 2015, Warsaw, Poland.}
}
\author{Radoslaw Ryblewski
\address{The H. Niewodnicza\'nski Institute of Nuclear Physics, Polish Academy of Sciences, PL-31342 Krak\'ow, Poland}
}
\maketitle
\begin{abstract}
The framework of leading-order anisotropic hydrodynamics is supplemented with realistic equation of state and self-consistent freeze-out prescription. The model is applied to central proton-nucleus collisions. The results are compared to those obtained within standard Israel-Stewart second-order viscous hydrodynamics. It is shown that the resulting hadron spectra are highly-sensitive to the hydrodynamic approach that has been used.
\end{abstract}
 %
\section{Introduction}
\label{sec:intro}
%
\par One of the most surprising outcomes of the ultra-relativistic heavy-ion experiments was to show that the new state of matter, the so called quark-gluon plasma (QGP), created in these reactions forms a strongly-coupled and correlated system with the smallest viscosity in Nature. As a result, its space-time evolution may be, to the great extent, described within the framework of relativistic viscous fluid dynamics. The application of the latter is complicated by the fact that at the early stages of the evolution there are locally large momentum-space anisotropies present in the system, which eventually make the standard framework to break down. In order to account for this a great theoretical development of the relativistic fluid dynamics was done in the recent years. In particular, a new fluid dynamical framework, called anisotropic hydrodynamics was proposed \cite{Martinez:2010sc,Florkowski:2010cf,Ryblewski:2010bs,
Martinez:2010sd,Ryblewski:2011aq,Martinez:2012tu,Ryblewski:2012rr,
Tinti:2013vba,Nopoush:2014pfa,Tinti:2014yya,Tinti:2015xwa}, which is based on the reorganization of the hydrodynamic expansion around the anisotropic background. At the same time, in order to account for small viscous-like corrections, the framework was generalized by adding linear perturbations to the leading-order term \cite{Bazow:2013ifa,Bazow:2015cha,Molnar:2016vvu}. Already at leading order the framework of anisotropic hydrodynamics was shown to agree very well with exact solutions of the underlying relaxation-type kinetic theory equations in the special cases of the Bjorken \cite{Florkowski:2013lza,Florkowski:2013lya,Florkowski:2014sfa} and Gubser \cite{Denicol:2014xca,Denicol:2014tha,Nopoush:2014qba} velocity profiles. Only recently the framework of anisotropic hydrodynamics was supplemented with the realistic quantum chromodynamics (QCD) equation of state and the freeze-out prescription allowing it to be applied to the central $A\textnormal{-}A$ and $p\textnormal{-}A$ collisions \cite{Nopoush:2015yga}, see also \cite{Alqahtani:2015qja}. In this proceedings contribution we briefly review the framework presented in Ref.~\cite{Nopoush:2015yga} and present its main results for $p\textnormal{-}A$ reactions.
%
\section{Leading-order anisotropic hydrodynamics}
\label{sec:intro}
%
In the case of central collisions at ultra-relativistic energies, considered in this work, one can restrict oneself to the boost-invariant and cylindrically symmetric (1+1)-dimensional system. Using the exact solutions of the kinetic theory equations \cite{Nopoush:2014qba} it was shown that within this symmetry the actual distribution function may be very well approximated by the ellipsoidal form $f^{\rm aniso}=f^{\rm eq}\left(\sqrt{p_\mu \Xi^{\mu\nu} p_\nu}/\lambda\right)$ \cite{Martinez:2012tu,Tinti:2013vba,Nopoush:2014qba}, where $f^{\rm eq}$ is an equilibrium distribution function, $\Xi^{\mu\nu} = u^\mu u^\nu + \xi^{\mu\nu}$, and the anisotropy tensor in local rest frame (LRF) reads  $\xi^{\mu\nu}_{\rm LRF}={\rm diag\left(0,\xi_x, \xi_y,\xi_z\right)}$ and satisfies $\xi^\mu_{\,\,\,\mu} =0$. The flow four-vector $u^\mu$ is restricted by the symmetry to the form $u^\mu=(\cosh \theta_\perp \cosh \eta,\sinh \theta_\perp \cos \phi,\sinh \theta_\perp \sin \phi,\cosh \theta_\perp \sinh \eta)$. The equations of motion of anisotropic hydrodynamics are obtained by taking first and second moments of the Boltzmann kinetic equation in the relaxation-time approximation \cite{Tinti:2013vba,Nopoush:2014pfa}. As a result one obtains four coupled partial differential equations for anisotropy parameters $\alpha_x$ and $\alpha_z$ ($\alpha_i =1/\sqrt{1+\xi_i}$), transverse momentum scale $\lambda$ and transverse rapidity of the fluid $\theta_\perp$ \cite{Tinti:2013vba,Nopoush:2014pfa}.
For a conformal system the following equation of state holds \cite{Nopoush:2014qba,Nopoush:2015yga}
\begin{eqnarray}
{\cal E} &=& {\cal E}_{\rm eq}(\lambda)\,{\cal R}(\alpha_x,\alpha_z) \, ,  \nonumber  \\
{\cal P}_x &=& {\cal P}_{\rm eq} (\lambda)\,{\cal H}_{Tx}(\alpha_x,\alpha_z) \, ,  \nonumber \\
{\cal P}_y &=& {\cal P}_{\rm eq}(\lambda)\,{\cal H}_{Ty}(\alpha_x,\alpha_z) \, ,  \nonumber \\
{\cal P}_z &=& {\cal P}_{\rm eq}(\lambda)\,{\cal H}_{L}(\alpha_x,\alpha_z) \, ,\nonumber  
\end{eqnarray}
where ${\cal E}_{\rm eq}$ and ${\cal P}_{\rm eq}$  are the equilibrium expressions for the temperature dependence of energy density and pressure of the ideal gas of massless particles, respectively, ${\cal P}_{i}$ are pressures acting in $i$-th direction and the functions ${\cal R}$ and  ${\cal H}$ are defined in Ref.~\cite{Nopoush:2015yga}. In order to include the realistic equation of state we exchange energy density and pressure of the ideal gas with the corresponding functions obtained within lattice QCD (lQCD) simulations by the Wuppertal-Budapest collaboration \cite{Borsanyi:2010cj} (${\cal E}_{\rm eq}\to{\cal E}_{\rm lQCD}$, ${\cal P}_{\rm eq}\to {\cal P}_{\rm lQCD}$). Another important ingredient required for a realistic description of the QGP evolution is the self-consistent description of the decoupling of the system. For this purpose we assume that the momentum distribution of hadron spectra results from the Cooper-Frye formula applied to the ellipsoidal form $f^{\rm aniso}$ on the isothermal freeze-out hypersurface. To set the initial energy density profile for our simulations we employ standard optical Glauber model.
%
\section{Results and conclusions}
%
\begin{figure}[t]
\centerline{%
\includegraphics[width=13cm]{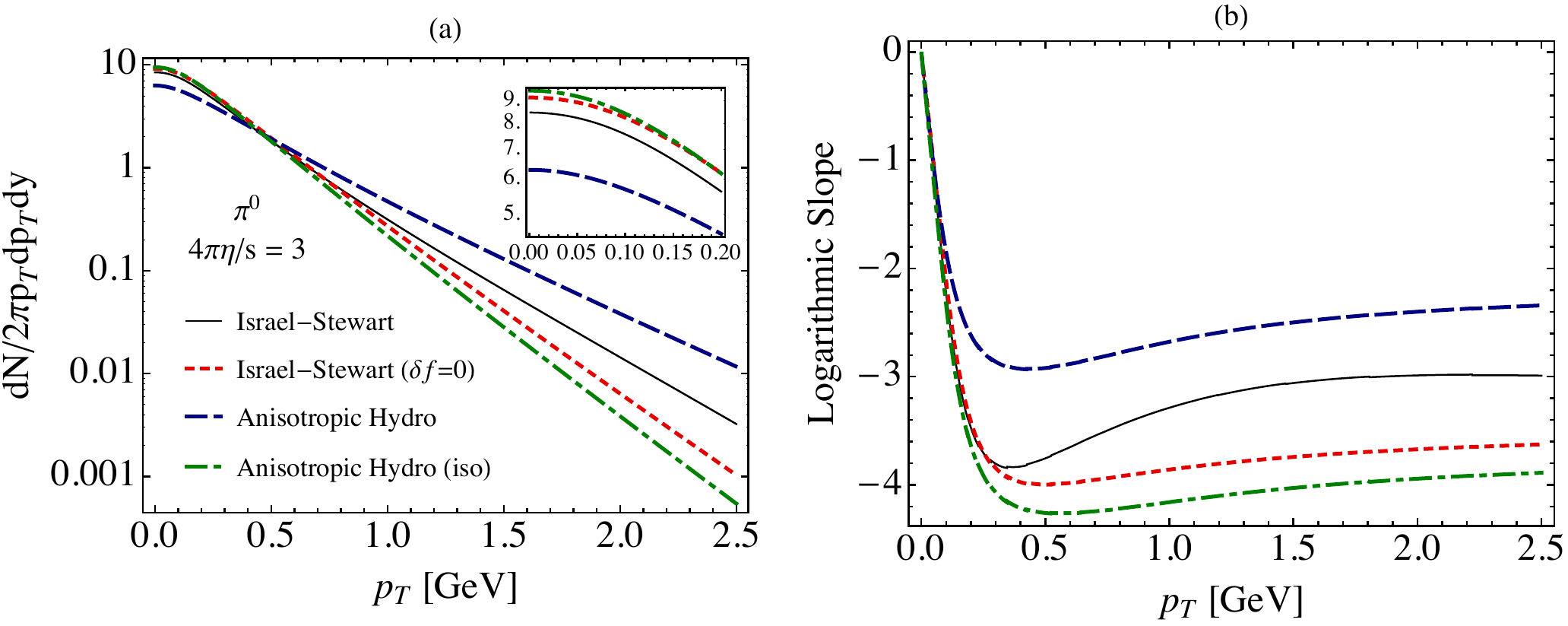}
}
\caption{(Color online) Transverse-momentum spectra of $\pi^0$ (a) and their logarithmic slope (b) for $p\textnormal{-}Pb$ collision obtained within leading-order anisotropic hydrodynamics and Israel-Stewart second-order viscous hydrodynamics.}
\label{fig:spectra}
\end{figure}
We apply the leading-order anisotropic hydrodynamics framework to the central $p\textnormal{-}Pb$ collisions, and compare its results with standard Israel-Stewart hydrodynamics. In Fig.~\ref{fig:spectra} we present the comparison of the resulting $\pi^0$ $p_T$-spectra (a) together with their logarithmic slope (b) for the two approaches. One can see that the former (dashed blue lines) predicts much harder spectra than the viscous hydrodynamics (solid black lines), which may be traced back to the fact that the pressure anisotropy is significantly overestimated in Israel-Stewart approach \cite{Nopoush:2015yga}. Moreover, one can also see that the self-consistent inclusion of anisotropy corrections at the freeze-out (dashed blue lines),  compared to the case where the corrections are neglected (dashed-dotted green lines), leads to large corrections to the final spectra. It means that in small systems the off-equilibrium corrections are significant at the time when the system decouples and must be included properly. 
\par Since the anisotropic hydrodynamics was shown to agree much better with the underlying kinetic theory than the Israel-Stewart theory  \cite{Florkowski:2013lza,Florkowski:2013lya,Florkowski:2014sfa,
Denicol:2014xca,Denicol:2014tha,Nopoush:2014qba} we conclude that the anisotropic hydrodynamics leads to dramatic improvement of the description of the evolution of matter in heavy-ion collisions, especially when one considers asymmetric $p\textnormal{-}Pb$ reactions.

\section*{Acknowledgments}
%
This contribution is dedicated to Prof. Jan Pluta on the occasion of his 70-th birthday. R.R. was supported by Polish National Science Center Grant No. DEC-2012/07/D/ST2/02125. 
%

\end{document}